\documentclass[12pt]{article}

\def\theorem #1. #2\par{\medbreak
  \noindent{\tt {\bf Theorem #1.}\enspace}{\sl#2\par}%
  \ifdim\lastskip<\medskipamount \removelastskip\penalty55\medskip\fi}

\def\for{\quad\hbox{for}\quad}
\def\R{{\cal R}}
\def\C{{\cal C}}
\def\K{{\cal K}}
\def\Tr{{\rm Tr}}
\def\diag{{\rm diag}}
\def\det{{\rm det}}
\def\Det{{\rm Det}}
\def\cliff{{\cal C}\!\ell_{p,q}^\K}
\def\Mat{{\rm Mat}}

\begin{document}
\baselineskip=18pt

\author{ N.G.Marchuk, S.E.Martynova}

\title{Notions of determinant, spectrum,
and Hermitian conjugation of Clifford
algebra elements}

\maketitle

nmarchuk@mi.ras.ru,   www.orc.ru/\~{}nmarchuk

martyno@yandex.ru

\begin{abstract}
We show how the matrix algebra notions of determinant, spectrum, and
Hermitian conjugation transfer to the Clifford algebra and to
differential forms on parallelisable manifolds.
\end{abstract}

{\noindent \bf Matrices.} Let $\Mat(n,\K)$ be the algebra of
$n$-dimensional matrices over the field $\K$ of real ($\K=\R$) or
complex ($\K=\C$) numbers. A matrix $P\in\Mat(n,\K)$ is said to be {\em
invertible} if there exists the matrix $P^{-1}\in\Mat(n,\K)$ such that
$PP^{-1}=P^{-1}P={\bf 1}$, where ${\bf 1}$ is the identity $n\times
n$-matrix. A number (scalar) $\lambda\in\C$ is called an {\em
eigenvalue} of a matrix $P\in\Mat(n,\K)$ if the matrix $P-\lambda{\bf
1}$ is not invertible. The full set of eigenvalues of a matrix $P$ is
called the {\em spectrum} of this matrix. The determinant of a matrix
$\det\,:\,\Mat(n,\K)\to\K$ is a standard notion of the linear algebra.
The determinant has the following properties:
\begin{itemize}
\item $\det(PQ)=\det(P) \det(Q)$, $P,Q\in\Mat(n,\K)$;
\item $\det(\alpha P)=\alpha^n \det(P)$, $\alpha\in\K$, $P\in\Mat(n,\K)$;
\item $\det({\bf 1})=1$;
\item A matrix $P\in\Mat(n,\K)$ is invertible iff $\det(P)\neq 0$.
\end{itemize}

It follows from the last property that all eigenvalues $\lambda$
of a matrix $P$
satisfy the equation
$$
\det(P-\lambda{\bf 1})=0.
$$
By $P^\dagger$ denote the Hermitian conjugated matrix (transposed matrix
with complex conjugated components). If $P^\dagger=P$, then the matrix
$P$ is called {\em Hermitian}. The operation of Hermitian conjugation of
matrices has the following properties:
\begin{itemize}
\item $(PQ)^\dagger=Q^\dagger P^\dagger$;
\item $P^{\dagger\dagger}=P$;
\item The spectrum of a Hermitian matrix is real.
\end{itemize}
The last property means that for any Hermitian matrix $P$ the equation
$\det(P-\lambda{\bf 1})=0$ has only real roots $\lambda$. In what follows
we show how the matrix algebra notions of determinant, spectrum, and
Hermitian conjugation transfer to the Clifford algebra $\cliff$.
\medskip

\noindent{\bf A Clifford algebra.}
Let $n$ be a natural number and let $L(n,\K)$ be
an $2^n$-dimensional linear space over the field $\K$ of real or complex
numbers. We suppose that the linear space $L(n,\K)$ has basis elements
enumerated by ordered multi-indices
$$
e, e^k, e^{k_1 k_2},\ldots,e^{12\ldots n}\quad
1\leq k\leq n,\,1\leq k_1<k_2<\cdots\leq n,\ldots.
$$
Elements of the linear space $L(n,\K)$ can be written in the form
\begin{equation}
U=ue+\sum_{1\leq k\leq n}u_k e^k+
\sum_{1\leq k_1<k_2\leq n}u_{k_1 k_2}e^{k_1 k_2}+\ldots +
u_{1\ldots n}e^{1\ldots n},
\label{Udecomp}
\end{equation}
where $u,u_k,\ldots,u_{1\ldots n}$ are scalars from the field $\K$.

Let $p+q=n$ and $\eta=\eta(p,q)\in\Mat(n,\R)$ be the diagonal matrix with $\pm1$ on
the diagonal
$$
\eta=\|\eta^{kl}\|=\diag(\underbrace{1,\ldots,1,}_{p-{\hbox{pieces}}}
\underbrace{-1,\ldots,-1}_{q-{\hbox{pieces}}}).
$$
Let us define a Clifford product $L(n,\K)\times L(n,\K)\to L(n,\K)$ by
the rules
\begin{itemize}
\item $L(n,\K)$ is an associative algebra with respect to the Clifford
product;
\item $e^k e^l+e^l e^k=2\eta^{kl}e$, $k,l=1,\ldots,n$;
\item $e^{k_1}\ldots e^{k_l}=e^{k_1\ldots k_l}$ for $k_1<\cdots<k_l$.
\end{itemize}
The $2^n$-dimensional linear space $L(n,\K)$ with the Clifford product
(defined with the aid of the matrix $\eta=\eta(p,q)$) is called the {\em
Clifford algebra} and is denoted by $\cliff$, where $p+q=n$. The
elements $e^1,\ldots e^n$ are said to be {\em generators} of $\cliff$.
We identify the unity element $e$ of $\cliff$ with the scalar unit $1$.

Denote by $\langle U\rangle_r$ the projection of an element $U\in\cliff$
on the linear subspace, which spans on the basis elements $e^{k_1\ldots
k_r}$, i.e., for $U$ from (\ref{Udecomp}) we have
$$
\langle U\rangle_r=\sum_{k_1<\cdots<k_r}u_{k_1\ldots k_r}e^{k_1\ldots
k_r}\quad\hbox{and}\quad
U=\sum_{r=0}^n \langle U\rangle_r.
$$

The projection $\Tr(U)=\langle U\rangle_0$ is called the {\em trace} of
$U\in\cliff$. It is easy to see that
$$
\Tr(UV-VU)=0,\quad \Tr(W^{-1}UW)=\Tr(U),
$$
where $U,V$ are arbitrary elements of $\cliff$ and $W$ is an arbitrary
invertible element of $\cliff$.

Now we consider four operations of conjugation in the Clifford algebra
$$
U^\wedge=U|_{e^k\to-e^k},\quad
U^\sim=U|_{e^{k_1\ldots k_r}\to e^{k_r}\ldots e^{k_1}},\quad
\bar{U}=U|_{u_{k_1\ldots k_r}\to\bar{u}_{k_1\ldots k_r}},
$$
where $\bar{u}_{k_1\ldots k_r}$ are complex conjugated scalars.  The
superposition of the operations $U^\sim$ and $\bar{U}$ gives the
opereation of {\em Clifford conjugation} $U^*=\bar{U}^\sim$.
Evidently,
\begin{eqnarray*}
U^\wedge&=&\langle U\rangle_0-\langle U\rangle_1+\langle U\rangle_2-
\langle U\rangle_3+\ldots,\\
U^\sim&=&\langle U\rangle_0+\langle U\rangle_1-\langle U\rangle_2-
\langle U\rangle_3+\ldots
\end{eqnarray*}
and
\begin{eqnarray*}
&&U^{\wedge\wedge}=U,\quad
U^{\sim\sim}=U,\quad
\bar{\bar{U}}=U,\quad
U^{**}=U,\\
&&(UV)^\wedge=U^\wedge V^\wedge,\,
(\overline{UV})=\bar{U}\bar{V},\,
(UV)^\sim=V^\sim U^\sim,\,
(UV)^*=V^* U^*.
\end{eqnarray*}
\medskip

\noindent{\bf Matrix representations of Clifford algebra elements.}
Let us take the Pauli matrices $\sigma^1,\sigma^2,\sigma^3$
$$
\sigma^0=\pmatrix{ 1 & 0\cr 0 & 1 },\quad
\sigma^1=\pmatrix{ 0 & 1\cr 1 & 0 },\quad
\sigma^2=\pmatrix{ 0 &-i\cr i & 0 },\quad
\sigma^3=\pmatrix{ 1 & 0\cr 0 &-1 }.\quad
$$
Consider the following matrices
$\underline{e}^1,\ldots,\underline{e}^n$, which are matrix
representation of generators $e^1,\ldots,e^n$ of the Clifford algebra
$\cliff$:
\begin{itemize}
\item $n=1$
$$
\underline{e}^1=\alpha^1 \sigma^1;
$$
\item $n=2$
$$
\underline{e}^1=\alpha^1 \sigma^1,\quad
\underline{e}^2=\alpha^2 \sigma^2;
$$
\item $n=3$
$$
\underline{e}^k=\alpha^k \pmatrix{\sigma^k & 0\cr 0 & -\sigma^k},
\quad k=1,2,3;
$$
\item $n=4$
$$
\underline{e}^4=\alpha^4\pmatrix{0 & \sigma^0\cr \sigma^0 & 0},\quad
\underline{e}^k=\alpha^k \pmatrix{\sigma^k & 0\cr 0 & -\sigma^k},
\quad k=1,2,3,
$$
\end{itemize}
where $\alpha^1=\cdots=\alpha^p=1$,
$\alpha^{p+1}=\cdots=\alpha^{p+q}=i$ and
$\underline{e}^k \underline{e}^l+\underline{e}^l\underline{e}^k=
2\eta^{kl}{\bf 1}$.
Hence, we may connect each element $U$ of $\cliff$ with the matrix
$\underline{U}$ such that
\begin{equation}
\underline{U}=U|_{e^k\to\underline{e}^k,\,1\to{\bf 1}},
\label{mat:rep}
\end{equation}
where ${\bf 1}$ is the identity matrix.
\medskip

\noindent{\bf Notions of determinant and spectrum of Clifford algebra
elements.}
Now we introduce a concept of determinant ($\Det$) of a Clifford algebra
element $U$. By definition, put
$$
\Det(U)= \det(\underline{U}).
$$
The proofs of the theorems A1,A2,A3 are by direct calculation.

\theorem A1. The determinant of $U\in\cliff$ satisfy the following
formulas:
$$
\Det(U)=\left\lbrace\begin{array}{ll}
\Omega_U& \for n=1,2;\\
\Omega^\sim_U\Omega_U & \for n=3;\\
\Tr(\Omega^\sim_U\Omega_U)
-2 \det(\eta)(\Tr(\ell\Omega_U))^2 &
\for n=4,
\end{array}
\right.
$$
where $\Omega_U=U^\wedge U^\sim$ and $\ell=e^{1234}$. It is easy to
check that $\Omega_U$ is a scalar for $n=1,2$ and
$\Omega^\sim_U\Omega_U$ is a scalar for $n=3$.
\par

\theorem A2. The determinant of Clifford algebra elements has the
properties
\begin{itemize}
\item $\Det(UV)=\Det(U)\Det(V)$ for $U,V\in\cliff$;
\item $\Det(\beta U)=\beta^{k(n)}\Det(U)$, where $\beta\in\K$,
$U\in\cliff$ and $k(n)=2$ for $n=1,2$ and $k(n)=4$ for $n=3,4$;
\item $\Det(1)=1$;
\item An element $U\in\cliff$ is invertible iff $\Det(U)\neq0$.
\end{itemize}
\par

If scalar $\lambda\in\C$ satisfy the equation $\Det(U-\lambda)=0$, then
we say that $\lambda$ is an eigenvalue of the element $U\in\cliff$. The
full set of eigenvalues of a Clifford algebra element $U$ is called the
spectrum of this element. The spectrum of a Clifford algebra element
consists of two scalars $\{\lambda_1,\lambda_2\}$ in cases $n=1,2$ and
consists of four scalars $\{\lambda_1,\ldots,\lambda_4\}$ in cases
$n=3,4$ (with regard to multiplicity).
\medskip

\noindent{\bf A Hermitian conjugation of Clifford algebra elements.}
Let a Clifford algebra element $U$ and matrix $\underline{U}$ be
connected by the formula (\ref{mat:rep}). Consider the Hermitian
conjugated matrix $(\underline{U})^\dagger$ and take the element
$V\in\cliff$ such that $\underline{V}=(\underline{U})^\dagger$. This
element is said to be {\em Hermitian conjugated} to the element $U$ and
is denoted by $V=U^\dagger$, i.e.,
\begin{equation}
V=U^\dagger\,\Longleftrightarrow \underline{V}=(\underline{U})^\dagger
\for U,V\in\cliff.
\label{herm:conj}
\end{equation}

\theorem A3. The Hermitian conjugated element $U^\dagger\in\cliff$ satisfy
the following formulas:
$$
U^\dagger=\left\lbrace\begin{array}{lll}
U^* & \for (p,q)=(n,0),&n=1,\ldots,4;\\
e^1 U^* e^1 & \for (p,q)=(1,n-1),&n=2,3,4;\\
-e^n U^{*\wedge} e^n & \for (p,q)=(n-1,1),&n=3,4;\\
-e^{12} U^{*\wedge} e^{12} & \for (p,q)=(2,2),&n=4;\\
U^{*\wedge} & \for (p,q)=(0,n),&n=1,\ldots,4.
\end{array}
\right.
$$
\par

An element $U\in\cliff$ is called Hermitian if $U=U^\dagger$. The
following theorem is a consequence of the formula (\ref{herm:conj}).

\theorem A4. Any Hermitian Clifford algebra element $U\in\cliff$ has real
spectrum.
\par

\medskip

\noindent{\bf Conclusions.}
Therefore using matrix representations we introduce notions of
determinant, spectrum, and Hermitian conjugation of Clifford algebra
elements. In theorems A1,A3 we establish formulas for determinant
and for Hermitian conjugation in terms of intrinsic Clifford algebra
operations. Now we may define the determinant and the Hermitian
conjugation of Clifford algebra elements using formulas of theorems
A1,A3 without reference to matrix representations.
\medskip

\noindent{\sl Definition 1.} The determinant of a Clifford algebra
element $U\in\cliff$ is the scalar $\Det(U)\in\K$ from the formula in
Theorem A1.
\medskip

\noindent{\sl Definition 2.} The operation of Hermitian conjugation of a
Clifford algebra element $\dagger:\cliff\to\cliff$ is the operation
defined by the formula in Theorem A3.
\medskip

In \cite{DTTE} we discuss a four dimensional parallelisable manifold
${\cal W}$ with the tetrad 1-forms $e^a$, $a=0,1,2,3$. These 1-forms can
be considered as generators of the Clifford algebra ${\cal
C}\!\ell_{1,3}$ and differential forms  on ${\cal W}$ can be considered
as elements of the Clifford algebra ${\cal
C}\!\ell_{1,3}$ with respect to the Clifford product of differential
forms \cite{DTTE}. This implies that the discussed Clifford algebra
notions of determinant, spectrum, and Hermitian conjugation are
applicable to differential forms on parallelisable manifolds.


\end{document}